\documentclass[10pt,nofootinbib,notitlepage,twocolumn,superscriptaddress]{revtex4-1}

\usepackage[utf8]{inputenc}
\usepackage[T1]{fontenc}
\usepackage[english]{babel}  

\usepackage{times}

\usepackage{amssymb,amsmath,amsfonts,amsthm}
\usepackage{mathtools}
\usepackage{verbatim}
\usepackage{enumerate}
\usepackage{graphicx}
\usepackage{hyperref}
\usepackage{bbm}
\usepackage{booktabs}

\usepackage[caption=false]{subfig}

\usepackage{color}
\definecolor{dred}{rgb}{.8,0.2,.2}
\definecolor{ddred}{rgb}{.8,0.5,.5}
\definecolor{dblue}{rgb}{.2,0.2,.8}
\definecolor{dgreen}{rgb}{.2,0.5,.2}

\theoremstyle{plain}

\theoremstyle{definition}

\newcommand{\bra}[1]{\mbox{$\langle #1|$}}
\newcommand{\ket}[1]{\ensuremath{|#1\rangle}}

\DeclareMathOperator{\diag}{diag}

\newcommand{\be}{\begin{equation}}
\newcommand{\ee}{\end{equation}}

\def\1#1{{\bf #1}}
\def\2#1{{\cal #1}}
\def\7#1{{\mathbb #1}}

\newcommand{\bea}{\begin{eqnarray}}
\newcommand{\eea}{\end{eqnarray}}

\newcommand{\op}[2]{|#1\rangle \langle #2|}
\newcommand{\tr}{\textrm{tr}}

\begin{document}

\title{Tomography is necessary for universal entanglement detection
with single-copy observables}

\author{Dawei Lu}
\thanks{These authors contributed equally to this work.}
\affiliation{Institute for Quantum Computing,
University of Waterloo, Waterloo N2L 3G1, Ontario, Canada}

\author{Tao Xin}
\thanks{These authors contributed equally to this work.}
\affiliation{Institute for Quantum Computing,
University of Waterloo, Waterloo N2L 3G1, Ontario, Canada}
\affiliation{State Key Laboratory of Low-Dimensional Quantum Physics and Department of Physics, Tsinghua University, Beijing 100084, China}

\author{Nengkun Yu}
\thanks{These authors contributed equally to this work.}
\affiliation{Institute for Quantum Computing,
University of Waterloo, Waterloo N2L 3G1, Ontario, Canada}
\affiliation{Department of Mathematics \& Statistics, University of
  Guelph, Guelph, Ontario, Canada}%

\author{Zhengfeng Ji}
\affiliation{Institute for Quantum Computing,
University of Waterloo, Waterloo N2L 3G1, Ontario, Canada}
\affiliation{State Key Laboratory of Computer Science, Institute of
  Software, Chinese Academy of Sciences, Beijing, China}%

\author{Jianxin Chen}
\address{Joint Center for Quantum Information and Computer Science,
  University of Maryland, College Park, Maryland, USA}

\author{Guilu Long}
\affiliation{State Key Laboratory of Low-Dimensional Quantum Physics and Department of Physics, Tsinghua University, Beijing 100084, China}

\author{Jonathan Baugh}
\affiliation{Institute for Quantum Computing,
University of Waterloo, Waterloo N2L 3G1, Ontario, Canada}

\author{Xinhua Peng}
\affiliation{Hefei National Laboratory for Physical Sciences at Microscale and Department of Modern Physics, University of Science
and Technology of China, Hefei, Anhui 230036, China}

\author{Bei Zeng}
\email{zengb@uoguelph.ca}
\affiliation{Institute for Quantum Computing,
University of Waterloo, Waterloo N2L 3G1, Ontario, Canada}
\affiliation{Department of Mathematics \& Statistics, University of
  Guelph, Guelph, Ontario, Canada}
\affiliation{Canadian Institute for Advanced Research, Toronto,
  Ontario, Canada}

\author{Raymond Laflamme}
\affiliation{Institute for Quantum Computing,
University of Waterloo, Waterloo N2L 3G1, Ontario, Canada}
\affiliation{Canadian Institute for Advanced Research, Toronto,
  Ontario, Canada}
\affiliation{Perimeter Institute for Theoretical Physics, Waterloo N2L 2Y5, Ontario,
Canada}%

\begin{abstract}
  Entanglement, one of the central mysteries of quantum mechanics,
  plays an essential role in numerous applications of quantum
  information theory. A natural question of both theoretical and
  experimental importance is whether universal entanglement detection
  is possible without full state tomography. In this work, we prove a
  no-go theorem that rules out this possibility for any non-adaptive
  schemes that employ single-copy measurements only. We also examine
  in detail a previously implemented experiment, which claimed to
  detect entanglement of two-qubit states via adaptive single-copy
  measurements without full state tomography. By performing the
  experiment and analyzing the data, we demonstrate that the
  information gathered is indeed sufficient to reconstruct the state.
  These results reveal a fundamental limit for single-copy
  measurements in entanglement detection, and provides a general
  framework to study the detection of other interesting properties of
  quantum states, such as the positivity of partial transpose and the
  $k$-symmetric extendibility.
\end{abstract}

\maketitle

\section*{Introduction}

Entanglement is one of the central mysteries of quantum mechanics---two or more
parties can be correlated in some way that is much stronger than they
can be in any classical way. Famous thought experiments questioning
the essence of quantum entanglement include the EPR
paradox~\cite{einstein1935can} and the Schrodinger's
cat~\cite{schrodinger1935present}, which ask the fundamental question
whether quantum mechanics is incomplete and there are hidden variables
not described in the theory. These debates about the weirdness of
quantum mechanics were later put into a theorem by
Bell~\cite{bell1964einstein}, which draws a clear line between
predictions of quantum mechanics and those of local hidden variable
theories. Bell's theorem was tested extensively in
experiments~\cite{clauser1978bell,aspect1981experimental,pan2000experimental,rowe2001experimental,groblacher2007experimental,ansmann2009violation,giustina2013bell,christensen2013detection,hensen2015experimental}
and quantum mechanics stands still to date.

More concretely, a bipartite quantum state $\rho_{AB}$ of systems $A$
and $B$ is separable if it can be written as a mixture of product
states $\rho_{AB} = \sum_i p_i \rho_A^i \otimes \rho_B^i$ with
$p_i\geq 0$ and $\sum_i p_i=1$, for some states $\rho_A^i$ of system
$A$ and $\rho_B^i$ of system $B$; otherwise, $\rho_{AB}$ is
entangled~\cite{werner1989quantum}. However, not every entangled state
$\rho_{AB}$ violates Bell inequalities---some entangled states
do allow local hidden variable descriptions~\cite{werner1989quantum}.

In practice, entanglement may also be detected by measuring the
`entanglement witnesses', physical observables with certain values
that prove the existence of quantum entanglement in a given state
$\rho_{AB}$~\cite{Horodecki2009}. However, none of these entanglement
witnesses could be universal. That is, the value of an entanglement
witness cannot tell with certainty whether an arbitrary state is
entangled or not. On the other hand, the `entanglement' measures do
play such a universal role. By commonly accepted axioms, the quantum
state $\rho_{AB}$ is entangled if and only if it has a nonzero value
of any entanglement measure~\cite{Plenio2007}. Unfortunately,
entanglement measures are not physical observables.

These commonly-known restrictions on Bell inequalities, entanglement
witnesses and entanglement measures raise a fundamental question: how
do we universally detect entanglement through physical observables?
The traditional approach to this problem is to completely characterize
the quantum state by means of state tomography~\cite{cramer2010efficient,hofheinz2009synthesizing,jullien2014quantum},
a method that provides
complete information of the state including, of course, entanglement
measures of the state. However, performing quantum state tomography
requires a large number of measurements, a daunting task for growing
system sizes.

A natural idea is to find a way to obtain the value of an entanglement
measure without FST. In fact, there have been
a lot of efforts along this line over the past
decade~\cite{guhne2009entanglement,mintert2007entanglement,horodecki2006general,horodecki2003measuring,augusiak2008universal,bartkiewicz2015method,horodecki2002method,carteret2005noiseless,guhne2007estimating,horodecki2003limits,bovino2005direct,huber2010detection,rudnicki2011collective,jungnitsch2011taming}.
However, common techniques to achieve this purpose rely heavily on
collective measurements on many identical copies of the state
$\rho_{AB}$. That is, joint measurement on more than one copy of the
state ($\rho_{AB}^{\otimes r}$ for some integer $r>1$) is needed. This
is bad news for experimentalists, as collective measurements are
usually much more difficult to implement than measuring single-copy
observables. It is then highly desirable to find a method that detects
entanglement without FST by measuring only
single-copy observables. The seeking of such a method has
been pursued in recent years with both theoretical simulations and
experimental realizations, leading to positive signs of
realizing such an appealing
task~\cite{park2010construction,van2012measuring}.

In this work, we examine the possibility of detecting entanglement
without FST by measuring only single-copy
observables. Surprisingly, despite the previous signs, we find that
this appealing task is unfortunately impossible, if only single-copy
observables are measured. That is, there is no way to
determine with certainty of any entanglement measure, or even to
determine whether the value is zero or not, without FST.
To be more precise, this means that for any set of
informationally-incomplete measurements, there always exists two
different states, an entangled $\rho_{AB}$ and a separable
$\sigma_{AB}$, giving the same measurement results under this
measurement. This sounds very counter-intuitive at the first sight, as
entanglement is just a single value, while quantum state tomography
requires measuring a set of observables that are
informationally-complete, scaling as the squared dimension
of the Hilbert space of the system.

Our observation is that universal detection of any property
without FST enforces strong geometrical structural conditions on the
set of states having that property. The set of separable states does
not satisfy such conditions due to its nonlinear nature and,
therefore, universal detection of entanglement without FST using
single-copy measurements is not possible. There is a nice geometric
picture of this fact: unless the shape of the separable states is
`cylinder-like', it is not possible to find a projection of the state
space to a lower dimensional hyperplane with non-overlapping image for
the set of separable states and entangled states.

If one allows adaptive measurements (the observable to be measured can
depend on previous measurement results), a protocol was implemented
in~\cite{park2010construction}, claiming to have detected entanglement
of a two-qubit state $\rho_{AB}$ via single-copy measurements without
FST. The protocol involves local filters that require repeated
tomography on each single qubit, which leads to a bound on the
entanglement measure concurrence~\cite{Wootters1998} of $\rho_{AB}$,
in case the single-qubit reduced density matrices $\rho_A$ and
$\rho_B$ are not maximally mixed.

We design an experiment to implement this adaptive protocol as proposed
in~\cite{park2010construction}, and show that for
certain $\rho_{AB}$, given the experimental
data collected, the state $\rho_{AB}$ is already completely determined. In
other words, once the concurrence of $\rho_{AB}$ is
determined, the protocol already leads to a FST of $\rho_{AB}$, i.e.
the protocol does not lead to the universal detection of entanglement
without FST.
This supplements
our no-go result with non-adaptive measurements.

Additionally, it is worthy emphasizing that to our best knowledge this is the first experimental realization of quantum filters (or equivalently, the amplitude-damping channel) via the ancilla-assisted approach. Compared to the optical platform which does not demand extra ancilla qubits to realize an amplitude damping channel~\cite{wang2006experimental,park2010construction,fisher2012optimal}, our approach is more general and can be extended to other systems straightforwardly.

We further show that, however, if one
allows joint measurements on $r$-copies (i.e. $\rho_{AB}^{\otimes r}$) even for $r=2$,
one can indeed find protocols that detects the entanglement of $\rho_{AB}$ without
FST. Therefore our no-go result
reveals a fundamental limit for single-copy measurements, and provides
a general framework to study the detection of other interesting
quantities for a bipartite quantum state, such as the positivity of
partial transpose~\cite{Horodecki1996} and $k$-symmetric
extendibility~\cite{Doherty2004}.

\section*{Results}

We discuss a no-go result stating that it is impossible to determine universally
whether a state is entangled or not without FST, with only
single-copy measurements. We first
prove a no-go theorem for non-adaptive measurements, and then
examine the protocol with adaptive measurements as proposed
in~\cite{park2010construction} in detail. We design
an experiment to implement this adaptive protocol, and
demonstrate that the information gathered is indeed sufficient
to reconstruct the state.

\textbf{Non-adaptive measurement.} For any given bipartite state $\rho_{AB}$,
one is only allowed to measure
physical observables on one copy of this given state. That is, we can
only measure Hermitian operators $S_k$ that are acting on
$\mathcal{H}_A \otimes \mathcal{H}_B$. For simplicity, we consider the
case where both $A$, $B$ are qubits. Our method
naturally extends to the general case of any bipartite systems (see
the Supplementary Information for details).

Now we consider a two-qubit state $\rho_{AB}$. In order to obtain some
information about $\rho$, we measure a set $\mathcal{S}$ of physical
observables $\mathcal{S}=\{S_1,S_2,\cdots,S_k\}$. An
informationally-complete set of observables contains $k=15$
linearly independent $S_i$'s. A simple choice of $\mathcal{S}$ is the
set of all two-qubit Pauli matrices other than the identity, i.e.
$\mathcal{S}=\{\sigma_i\otimes\sigma_j\}$ with $i,j=0,1,2,3$, where
$\sigma_0=I, \sigma_1=X, \sigma_2=Y, \sigma_3=Z$ and $(i,j)\ne (0,0)$.

Assume that we can decide universally whether an arbitrary $\rho_{AB}$
is entangled or not, without measuring an
informationally-complete set of observables. That is, there
exists a set $\mathcal{S}$ of at most $k=14$ physical observables such
that, by measuring $\mathcal{S}$, we can tell for sure whether
$\rho_{AB}$ is entangled or not. For our purpose, it suffices to
assume $k=14$.

The set of all two-qubit state $\rho_{AB}$, denoted as $\mathcal{A}$,
is characterized by $15$ real parameters, forming a convex set in
$\mathbb{R}^{15}$. The separable two-qubit states $\mathcal{S}$ form a
convex subset of $\mathcal{A}$. It is well-known that $\mathcal{S}$
has a non-vanishing volume~\cite{Zyczkowski1998}. Denote the set of
entangled states by $\mathcal{E}$, i.e.,
$\mathcal{E} = \mathcal{A} \setminus \mathcal{S}$.

\begin{figure}[htb]
  \label{fig:main}
  \centering
  \includegraphics[width=\columnwidth]{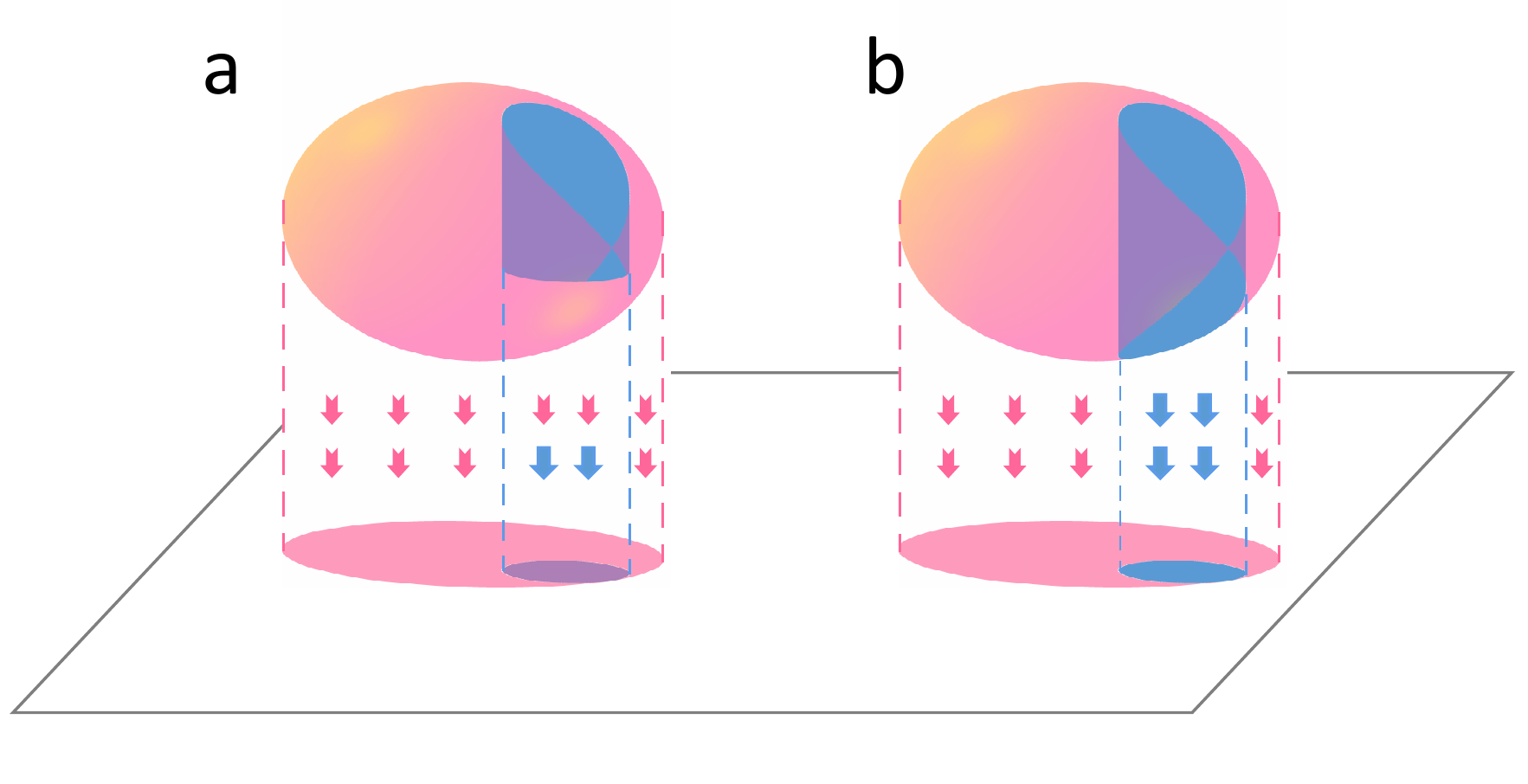}
  \setlength{\abovecaptionskip}{-0.00cm}
  \caption{\footnotesize{\textbf{Geometry of separable and entangled states.} The top pink oval represents
      the set of all states, denoted by $\mathcal{A}$.  Figure (b) shows a set (indicated by blue) that
      is an intersection of a generalized cylinder with $\mathcal{A}$ (i.e.
      `cylinder-like').
      The projection onto the plane that is orthogonal to the boundary
      lines of the cylinder separates this set with the rest of the states.
      Figure (a) has a set (indicated by blue) inside $\mathcal{A}$ which is not
      `cylinder-like'. Hence in fact no projection onto any plane exists
      that can separable the set with the rest states. The bottom
      ovals are the images of the top sets onto a plane, which clear
      show a separation of the images of the blue set from the pink set in Figure (b),
      but an overlap of images in Figure (a).}}  \label{fig:theory}
\end{figure}

The set of measurements $\mathcal{S}$ with $k=14$ can be visualized as
the definition of projections of $\mathcal{A}$ (hence also
$\mathcal{S}$) onto a $14$-dimensional hyperplane. If the measurement
of observables in $\mathcal{S}$ can tell for sure whether $\rho_{AB}$
is entangled or not, the images on the hyperplane of the separable
states $\mathcal{S}$ and the entangled states $\mathcal{E}$ must have
no overlap. We illustrate this geometric idea in
Fig.~\ref{fig:theory}.

In fact, the only possibility to separate any set from the rest of the
states without FST is that the set is an intersection
of the set of all states (i.e. set $\mathcal{A}$ as in Fig.~\ref{fig:theory})
with a generalized cylinder (i.e. a set of the form $\Omega\times (-\infty,+\infty)$,
where $\Omega$ is a convex set of dimension $14$),  In this sense,
we call these sets `cylinder-like', where the corresponding states can be
separated from the rest of states from some $14$ (or lower) dimensional projection.

Hence, to show that entanglement detection
without tomography is impossible, it suffices to prove that
$\mathcal{S}$ is not `cylinder-like' (in $\mathbb{R}^{15}$). To do
this, we show that for any projection onto a $14$-dimensional
hyperplane with normal direction $R$, there always exists
a two-qubit state $\rho$ that is on the boundary of the set
$\mathcal{S}$, such that $\rho+tR$ is entangled for some $t$ (see
Supplementary Information for details). That is, $\rho$ and $\rho+tR$ have the same
image on the $14$-dimensional hyperplane.

This geometric picture leads to a general framework to study the
detection of other interesting quantities for a bipartite quantum
state with single-copy measurements. Indeed, our proof also showed
that the sets of states with positive partial transpose (PPT) is not
`cylinder-like', hence cannot be universally detected by single-copy
measurements without full state tomography. With a similar method, we can
show that the sets of states allowing $k$-symmetric extension are also
not `cylinder like', even for two-qubit system. This reveals a
fundament limit of single-copy measurements, that is, full state
tomography is essentially needed to universally detect many
non-trivial properties of quantum states (e.g., separability, PPT,
$k$-symmetric extendability, see
Supplementary Information for details).

\textbf{Adaptive measurement.}
In case of adaptive protocols,
the observable to be measured in each step can depend on previous
measurement results. This kind of measurement protocol can be
formulated as follows. First an observable $H_1$ is chosen, and
$\tr(H_1\rho)$ is measured. Suppose the measurement result is
$\alpha_1$. Based on $\alpha_1$, observable $H_{2,\alpha_1}$ is
chosen, and $\tr(H_{2,\alpha_1}\rho)$ is measured. Suppose the
measurement result is $\alpha_2$. Based on $\alpha_1,\alpha_2$,
observable $H_{3,\alpha_1,\alpha_2}$ is chosen,
$\tr(H_{3,\alpha_1,\alpha_2}\rho)$ is measured and so on.

The protocol in~\cite{park2010construction} to determine the
concurrence~\cite{Wootters1998} of a two-qubit state without
FST falls into the category of adaptive measurements. We
implement this protocol and show that given the
experimental data collected for certain state $\rho_{AB}$, this
protocol in fact leads to FST of $\rho_{AB}$. That is,
this protocol does not
lead to universal detection of entanglement without FST.

\begin{figure}[htb]
\begin{center}
\includegraphics[width= \columnwidth]{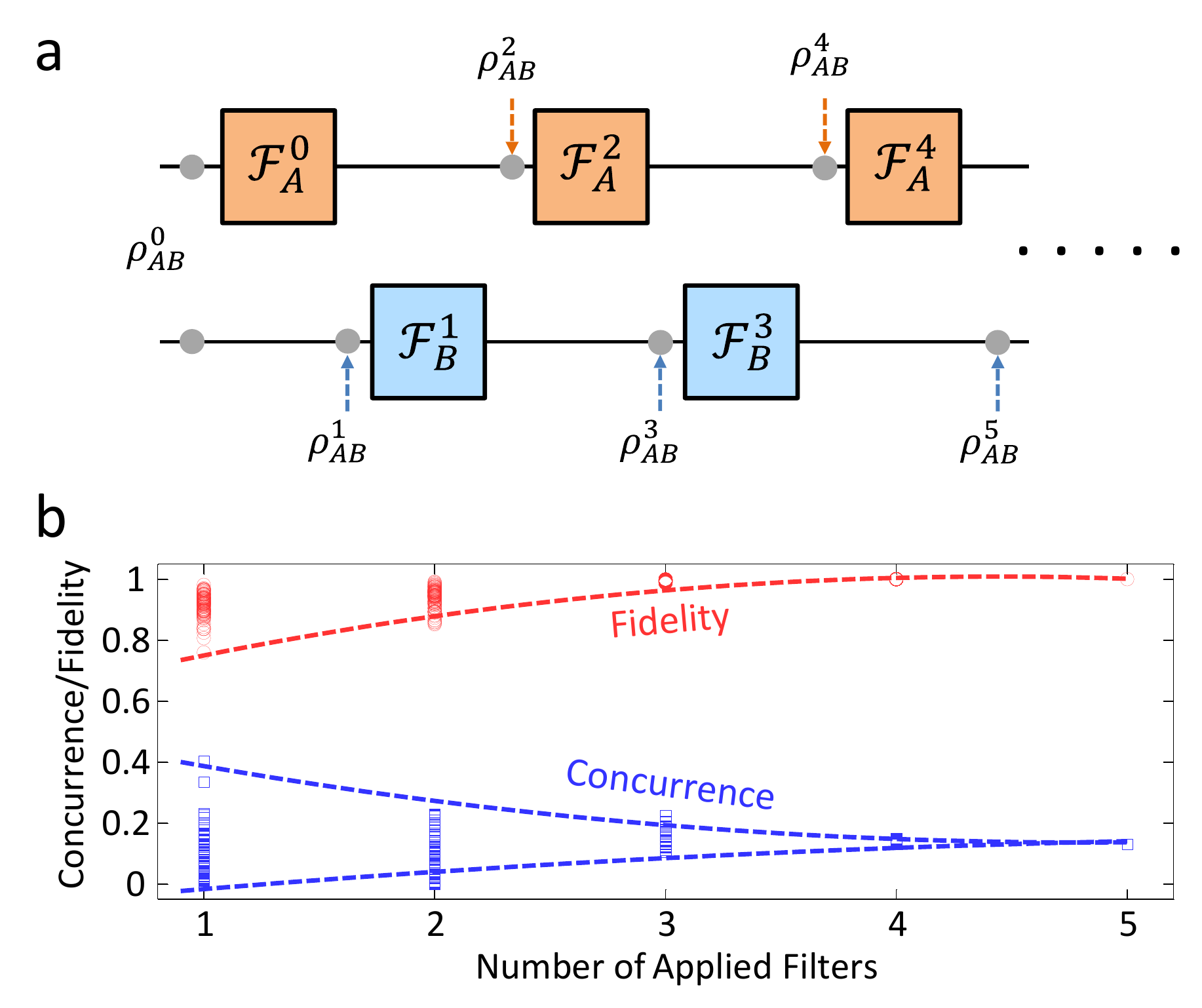}
\end{center}
\setlength{\abovecaptionskip}{-0.00cm}
\caption{\footnotesize{\textbf{Circuit and simulation results of the theoretical protocol.} (a) Schematic circuit for
    implementing the filter-based entanglement distillation proposal
    for an unknown two-qubit state $\rho_{AB}^{0}$.
    $\mathcal{F}_{A,B}^i=1/\sqrt{2\rho_{A,B}^{i}}$ ($i\geq 0$) is the
    $i$th local filter applied on A or B, where
    $\rho_{A}^{i}=\text{tr}_B(\rho_{AB}^{i})$ and
    $\rho_{B}^{i}=\text{tr}_A(\rho_{AB}^{i})$ are the local reduced
    density matrices of the current two-qubit state $\rho_{AB}^{i}$. The
    gray dots mean a single-qubit tomography is implemented at that
    place. (b) Simulated variation of concurrence and fidelity as the
    increase number $1\leq m \leq 5$ of applied filters. The simulated
    state is chosen as equation ~\eqref{inputstate} with $\lambda=0.2$. For
    any given $m$, we collected all the available reduced density
    matrices at this stage and reconstructed 100 possible input state.
    When $m \leq 4$, the reconstructed state is not unique due to the
    lack of constraints, so both the concurrence and fidelity have
    some distributions. When $m=5$, the input state can be uniquely
    determined, and the concurrence and fidelity converges to a single
    point. The dashed blue and red curves show the envelopes of the
    variations of concurrence and fidelity along with $m$,
    respectively. }}
\label{scheme}
\end{figure}

First let us briefly introduce the idea of entanglement distillation
via an iteratively filtering procedure~\cite{park2010construction}
depicted in Fig.~\ref{scheme}a. For an unknown two-qubit state
$\rho_{AB}^{0}$, we measure the local reduced density matrices
$\rho_{A}^{0}=\tr_B(\rho_{AB}^{0})$ and
$\rho_{B}^{0}=\tr_A(\rho_{AB}^{0})$ for both qubits. In case
$\rho_{A}^{0}$ and $\rho_{B}^{0}$ are not fully mixed, we
design the first filter $\mathcal{F}_A^0 = 1/\sqrt{2\rho_{A}^{0}}$
based on the information of $\rho_{A}^{0}$, and evolve $\rho_{AB}^{0}$
to $\rho_{AB}^{1}$ by applying $\mathcal{F}_A^0$. Similarly, the same
procedure is repeated for qubit B. The iterative applications of
filters are kept on going, and at step $k$,
the reduced density
matrices of the qubits will be $\rho_{A}^{k}$ and $\rho_{B}^{k}$.

In case both $\rho_A^{0}$
and $\rho_B^{0}$ are not identity, the iterative procedure described
above leads to a `distillation' of the density matrices
 $\rho_{A}^{k}$ and $\rho_{B}^{k}$ and it is guaranteed
 that they both converge to identity eventually~\cite{verstraete2003normal}.
All of the reduced density matrices  $\rho_{A}^{i}$ and $\rho_{B}^{i}$ ($i=0,1,\ldots,k$) are recorded
during the iterative procedure. At step $k$, when $\rho_{A}^{k}$ and $\rho_{B}^{k}$
are sufficient close to identity,
they can be used to reconstruct
a bound on the value of entanglement in $\rho_{AB}^{0}$
through the optimal witness $W(\rho_{AB}^{0})$ that is
only dependent on $\rho_{A}^{i}$ and $\rho_{B}^{i}$ ($i=0,1,\ldots,k$) (up to
local unitary transformations),
whose value hence tells whether  $\rho_{AB}^{0}$
is entangled or not~\cite{park2010construction}.

At the first sight, the above procedure seems feasible to determine
the value of entanglement without FST, since only single-qubit density matrices
$\rho_{A}^{i}$ and $\rho_{B}^{i}$ ($i=0,1,\ldots,k$) are repeatedly measured
and only local unitary transformations are used in constructing the optimal witness.
That is, it seems that the two-qubit correlations in $\rho_{AB}^{0}$ are never
measured, which hence not lead to FST.
However, a detailed look shows it is not the case.
The key observation here is that, `local filters' are in fact
`weak' measurements that do record the correlations in $\rho_{AB}^{0}$.
This is because that the filters cannot be implemented with probability one,
so the correlation in $\rho_{AB}^{0}$
is `encoded' in the information that all the filters are implemented successfully. In other
words, what these local filters and local tomography on each single qubit
does, is in fact an FST of $\rho_{AB}^{0}$.

In order to demonstrate the relationship between
the local filters and FST, we simulate the local filter procedure
by choosing different number of applied filters as depicted in
Fig.~\ref{scheme}a. It turns out, in many case
$k=4$ (five filters) is enough to uniquely determine
$\rho_{AB}^{0}$ based on the data of
$\rho_{A}^{i}$ and $\rho_{B}^{i}$ ($i=0,1,\ldots,k$). Thus,
the information of $\rho_{A}^{i}$ and $\rho_{B}^{i}$ lead
to an FST of $\rho_{AB}^{0}$.

As an example, we illustrate the simulation with the input state chosen as equation (\ref{inputstate}) with $\lambda=0.2$,  and the result is shown in
Fig.~\ref{scheme}b.
Initially, we have $15$ real parameters (i.e. degrees of freedom, DOF for short) to determine $\rho_{AB}^{0}$
(ignoring the identity part due to the normalization condition). When more and more
filters are applied, DOF is decreasing eventually since we are acquiring more and more knowledge about the original input state. For example, the initial local reduced density matrices $\rho_{A}^{0}$ and $\rho_{B}^{0}$ before applying any filter can already reduce DOF to $9$; $\rho_{B}^{1}$ after the first filter provides three more constraints so DOF lowers to $6$, and so on. It is found that with $5$ filters, the input
state $\rho_{AB}^{0}$ can be uniquely determined via the collected
information of the reduced density matrices. And this
procedure works similarly for many other two-qubit state $\rho_{AB}^{0}$,
where $5$ filters are found to be enough to reconstruct $\rho_{AB}^{0}$,
as we will show in our experiment results.

\textbf{Experimental protocol in NMR setup.} To experimentally implement the protocol as presented in Fig.~\ref{scheme}a,
we first discuss how to realize the local filters in NMR system.
Without loss of generality, we can consider a local filter
$\mathcal{F}_A$ applied on qubit A as an example. For any
$\mathcal{F}_A$, it can always be decomposed into the form of
$U_A\Lambda_A V_A$ via singular value decomposition, where $U_A$ and
$V_A$ are single-qubit unitaries and $\Lambda_A$ is a diagonal Kraus
operator
\begin{equation}
\Lambda_A =
\left(
  \begin{array}{cc}
    1 & 0 \\0 & \sqrt{ 1-\gamma_A}\\
  \end{array}
\right).
\end{equation}
$\gamma_A\in [0,1]$ relies on $\mathcal{F}_A$ and indicates the
probability that the excited state $\ket{1}$ decays to the ground
state $\ket{0}$ when a system undergoes $\Lambda_A$. Although
non-unitary, $\Lambda_A$ can be expanded to a two-qubit unitary with
the aid of an ancilla qubit 1.  Basically, if a two-qubit unitary can
transform
\begin{equation}
  \label{2qubittransform}
  \begin{split}
    \ket{0}_1\ket{0}_A &\rightarrow \ket{0}_1\ket{0}_A,\\
    \ket{0}_1\ket{1}_A &\rightarrow \sqrt{1-\gamma_A}\ket{0}_1\ket{1}_A+\sqrt{\gamma_A}\ket{1}_1\ket{1}_A,
  \end{split}
\end{equation}
the quantum channel on the system qubit A would be $\Lambda_A$ by
post-selecting the subspace in which the ancilla qubit 1 is $\ket{0}$.
One possible unitary transformation that satisfies
equation ~\eqref{2qubittransform} is
\begin{equation}
\mathcal{U}_{1A}=\left(
  \begin{array}{cccc}
    1 & 0 & 0 & 0\\
    0 & \sqrt{1-\gamma_A} & 0 & \sqrt{\gamma_A}\\
    0 & 0 & 1 & 0 \\
    0 &  -\sqrt{\gamma_A} & 0 &  \sqrt{1-\gamma_A} \\
  \end{array}
\right).\label{U1A}
\end{equation}

The operation $\mathcal{U}_{1A}$ is thus a controlled rotation: when
the system qubit $A$ is $\ket{0}$, the ancilla remains invariant; when
$A$ is $\ket{1}$, the ancilla undergoes a rotation
$R_{-y}(\theta_A)=e^{i\theta_A\sigma_y/2}$ where
$\theta_A=2\text{arccos}\sqrt{1-\gamma_A}$. Therefore, in an
ancilla-assisted system with the ancilla initialized to $\ket{0}$,
the local filter $\mathcal{F}_A$ can be accomplished through a
two-qubit unitary gate $(I\otimes U_A)\mathcal{U}_{1A}(I\otimes V_A)$
followed by post-selecting the subspace in which ancilla is $\ket{0}$.

\begin{figure*}[htb]
\begin{center}
\includegraphics[width= 1.5\columnwidth]{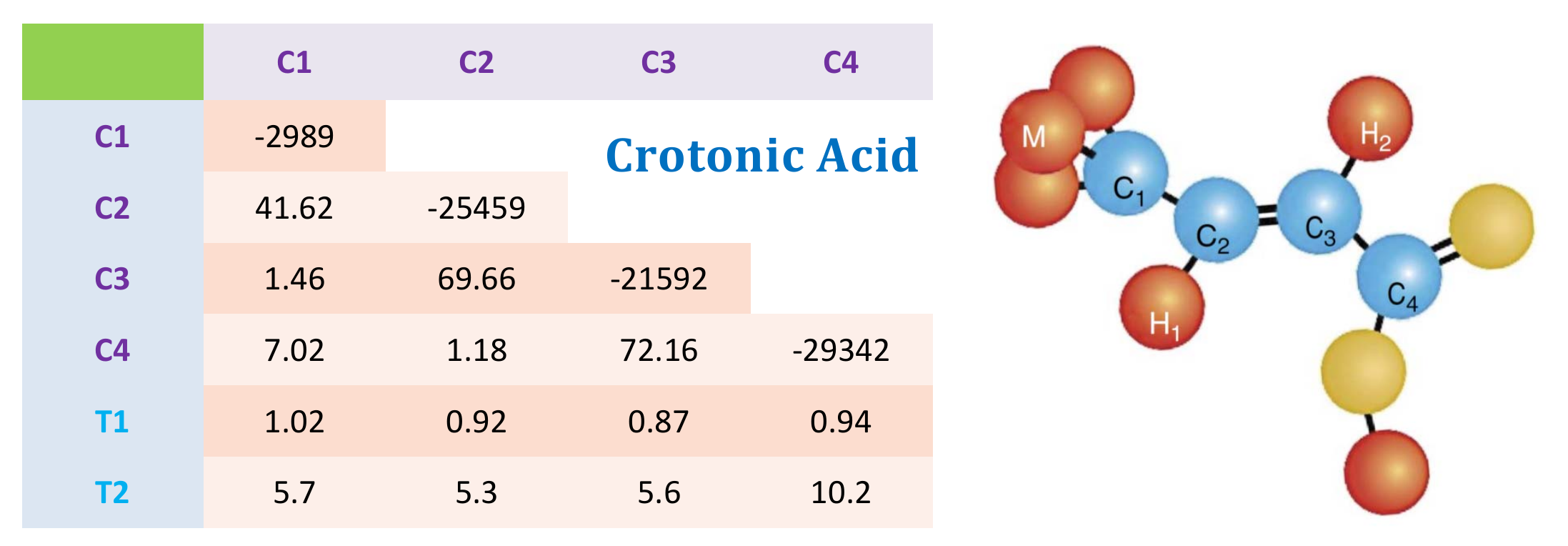}
\end{center}
\setlength{\abovecaptionskip}{-0.00cm}
\caption{\footnotesize{\textbf{Molecular structure and Hamiltonian parameters of $^{13}$C-labeled trans-crotonic acid.} C$_1$, C$_2$, C$_3$ and C$_4$ are used as four qubits in the experiment, and M, H$_1$ and H$_2$ are decoupled throughout the experiment. In the table, the chemical shifts with respect to the Larmor frequency 176.05MHz and J-coupling constants (in Hz) are listed by the diagonal and off-diagonal numbers, respectively. The relaxation time scales T$_{1}$ and T$_{2}$ (in Seconds) are shown at bottom.}}\label{molecule}
\end{figure*}

\textbf{NMR implementation.} To implement the aforementioned filter-based  entanglement distillation protocol as presented in Fig.~\ref{scheme}a in NMR, we need a 4-qubit quantum processor consisting of two system qubits A and B, and two ancilla qubits 1 and 2. Our 4-qubit sample is $^{13}$C-labeled trans-crotonic acid dissolved in d6-acetone. The structure of the molecule is shown in Fig.~\ref{molecule}, where C$_1$ to C$_4$ denote the four qubits. The methyl group M, H$_1$ and H$_2$ were decoupled throughout all experiments. The internal Hamiltonian of this system can be described as
\begin{align}\label{Hamiltonian}
\mathcal{H}_{int}=\sum\limits_{j=1}^4 {\pi \nu _j } \sigma_z^j  + \sum\limits_{j < k,=1}^4 {\frac{\pi}{2}} J_{jk} \sigma_z^j \sigma_z^k,
\end{align}
where $\nu_j$ is the chemical shift of the \emph{j}th spin and $\emph{J}_{jk}$ is the J-coupling strength between spins \emph{j} and \emph{k}. We assigned C$_3$ and C$_2$ as system qubits A and B, and C$_4$ and C$_1$ as ancilla qubits 1 and 2 to assist in mimicking the filters, respectively. All experiments were conducted on a Bruker DRX 700MHz spectrometer at room temperature.

\begin{figure*}[htb]
\begin{center}
\includegraphics[width= 2\columnwidth]{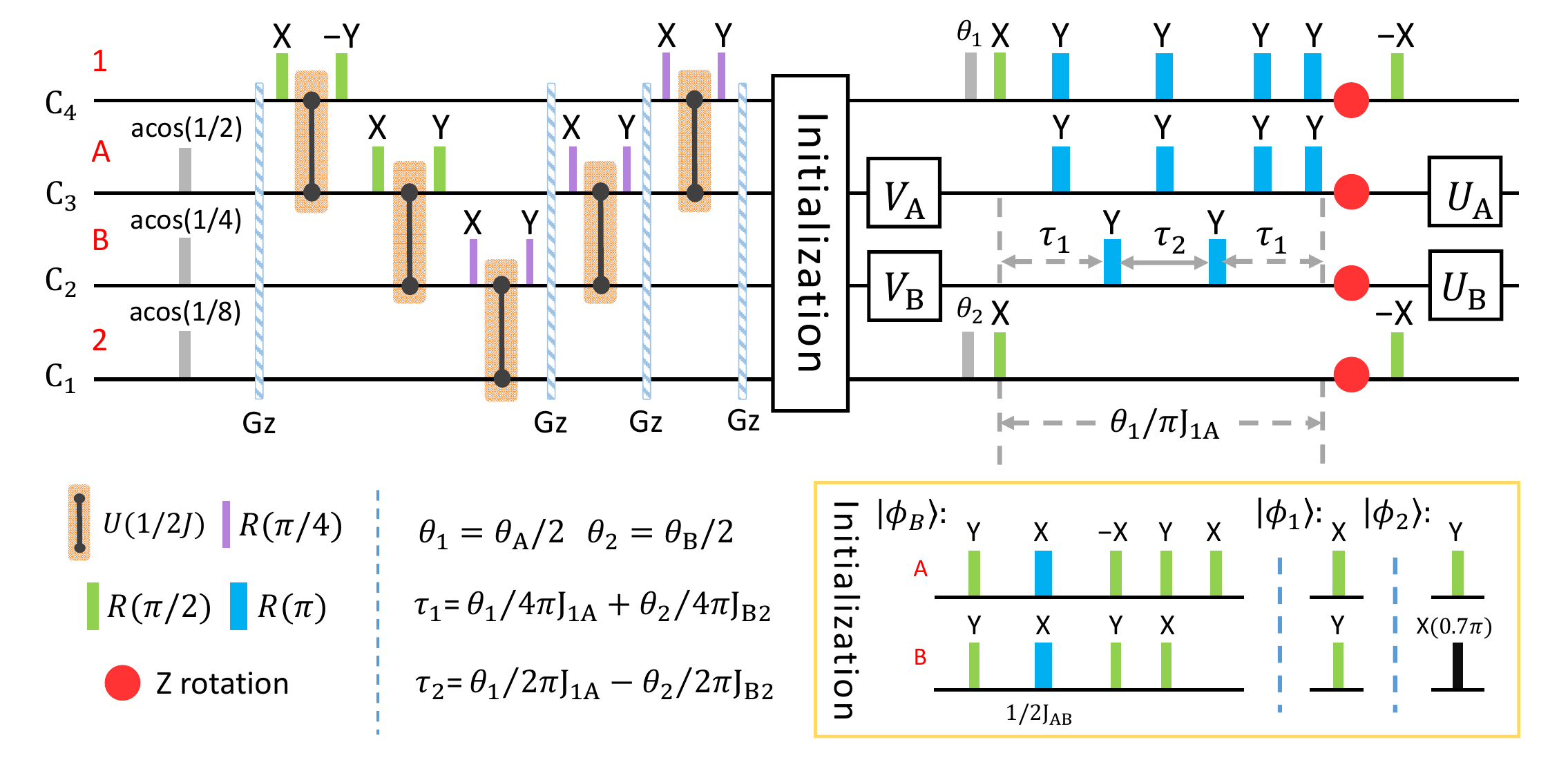}
\end{center}
\setlength{\abovecaptionskip}{-0.00cm}
\caption{\footnotesize{\textbf{NMR sequence to realize the filter-based proposal of entanglement distillation.} In particular, this sequence displays how to realize the first two filters $\mathcal{F}_{A}^0$ and $\mathcal{F}_{B}^1$ in terms of NMR pulses. All other sequences can be obtained analogously. A (marked by C$_3$) and B (marked by C$_2$) are system qubits to implement the proposal, while qubit 1 (marked by C$_4$) and 2 (marked by C$_1$) are ancilla qubits to assist in mimicking the filters. First the 4-qubit system is prepared to the PPS by spatial average technique which is shown before the initialization step. Then the system qubits are initialized to $\ket{\phi_{\mathcal{B}}}$,  $\ket{\phi_{1}}$ and $\ket{\phi_{2}}$ via three independent experiments illustrated in the lower-right inset, respectively.  The part after the initialization step is the sequence for realizing filters. $V_A$, $V_B$, $U_A$, $U_B$, $\theta_1$ and $\theta_2$ all depend on the measured results of reduced density matrices. Refer to the main text for detailed information of the parameters. }}\label{circuit}
\end{figure*}

Our target input state was chosen as a mixed state involving one Bell-state portion and two product-state portions, with the weight of Bell-state portion tunable. The state is written as
\be
\rho_{AB}^{0} = \lambda\ket{\phi_{\mathcal{B}}}\bra{\phi_{\mathcal{B}}}+(1-\lambda)\left ( \ket{\phi_{1}}\bra{\phi_{1}} + \ket{\phi_{2}}\bra{\phi_{2}}\right )/2,
\label{inputstate}
\ee
where
\begin{align}
\ket{\phi_{\mathcal{B}}} &= \left ( \ket{00} +\ket{11}\right )/\sqrt{2}, \\ \nonumber
\ket{\phi_{1}} &= \left ( \ket{0} -i\ket{1}\right )\left ( \ket{0} +\ket{1}\right )/2,\\ \nonumber
\ket{\phi_{2}} &= \left ( \ket{0} +\ket{1}\right )\left ( \ket{0} -2i\ket{1}\right )/\sqrt{10}
\end{align}
have concurrences $1$, $0$ and $0$, respectively. The parameter
$\lambda$ in $[0, 1]$ is thus proportional to the value of entanglement of
$\rho_{AB}^{0} $. In experiment, we varied $\lambda$ from 0.2 to 0.7
with step size 0.1 for every point, and implemented the proposal
correspondingly. Considering the two ancilla, the overall input state
for our 4-qubit system is thus
$\ket{0}\bra{0}\otimes\rho_{AB}^{0} \otimes\ket{0}\bra{0}$. As shown
in Fig.~\ref{circuit}, we prepared a pseudo-pure state from the
thermal equilibrium via spatial average technique~\cite{Cory04031997, dawei2011, 1367-2630-16-5-053015} and then created
three components $\ket{\phi_{\mathcal{B}}}$, $\ket{\phi_{1}}$ and
$\ket{\phi_{2}}$ on the system qubits, respectively (see Methods for detailed descriptions). Subsequently, each component undergoes the whole
filtering and single-qubit readout stage, with the final result
obtained by summarizing over all three experiments.

A two-qubit state tomography was implemented on the system qubits after creating $\rho_{AB}^{0}$. $\rho_0^{e}$ was reconstructed in experiment and its fidelity compared with the expected $\rho_{AB}^{0}$ is over 98.2\% for any $\lambda$ (Supplementary Table S1). This two-qubit state tomography is not required in the original proposal~\cite{park2010construction} in which only single-qubit measurements are necessary. However, since we claim that the filter-based proposal has already provided sufficient information to reconstruct the initial two-qubit state $\rho_0^{e}$, we need to compare it with $\rho_{f}^{e}$ which is reconstructed after running the entire proposal. To support our viewpoint, we have to show that $\rho_0^{e}$ and $\rho_{f}^{e}$ are the same up to minor experimental errors. This comparison is the only purpose of doing a two-qubit state tomography here.

Now we show how to realize local filtering operations in NMR. By measuring the local reduced density matrix $\rho_{A}^{0}$ of the input state $\rho_{AB}^{0}$, the first filter in Fig. \ref{scheme}a was calculated via $\mathcal{F}_A^0 = 1/\sqrt{2\rho_{A}^{0}}$ and decomposed into $U_A^0\Lambda_A^0 V_A^0$. Since $U_A^0$ and $V_A^0$ are merely local unitaries on qubit A, they can be realized by local radio-frequency (RF) pulses straightforwardly. $\Lambda_A^0$, which can be expanded to a 2-qubit controlled rotation  $\mathcal{U}_{1A}$ (see equation (\ref{U1A})) in a larger Hilbert space, was performed by a combination of local RF pulses and J-coupling evolutions\cite{TaoPRA}
\be
\mathcal{U}_{1A}=R_{-x}^1(\pi/2)U(\theta_A/2\pi J_{1A})R_{x}^1(\pi/2)R_{-y}^1(\theta_A/2),
\ee
where $U(\theta_A/2\pi J_{1A})$ represents the J-coupling evolution
$e^{-i\theta_A\sigma_z^1\sigma_z^A/4}$ between qubit 1 and A, and
$\theta_A=2\text{arccos}\sqrt{1-\gamma_A}$ depends on $\Lambda_A^0$.
After this filter, the system evolved to $\rho_{AB}^{1}$ and a
single-qubit tomography on qubit B was implemented, as shown by the
gray dots in Fig. \ref{scheme}a. The same procedure was repeated for
qubit B to realize the second filter  $\mathcal{F}_B^1 =
1/\sqrt{2\rho_{B}^{1}}$. In experiment, these two filters
$\mathcal{F}_A^0$ and $\mathcal{F}_B^1$ were carried out
simultaneously using the partial decoupling technique as shown in Fig.
\ref{circuit}, with additional Z rotations in the tail to compensate
the unwanted phases induced by the chemical shift evolutions. In Fig.
\ref{circuit}, $\theta_1$ and $\theta_2$ pulses are used to realize
$R_{-y}^1(\theta_A/2)$ and $R_{-y}^2(\theta_B/2)$, respectively, and
the free evolution time $\tau_1$ and $\tau_2$ are defined as
\bea
\tau_1 &=& \theta_1/4\pi J_{1A} + \theta_2/4\pi J_{B2} \\ \nonumber
\tau_2 &=& \theta_1/2\pi J_{1A} - \theta_2/2\pi J_{B2}.
\eea
Here we have assumed that $\tau_2>0$ ($\theta_1/2\pi J_{1A} > \theta_2/2\pi J_{B2}$). When $\tau_2<0$, the circuit just needs to be modified slightly by adjusting the positions of refocusing $\pi$ pulses. All the other filters have analogical structures with the one shown in Fig.~\ref{circuit}, and they were always carried out on qubit A and B simultaneously since they commute.

Every time after performing one local filter, we implemented a single-qubit tomography on the other qubit rather than the working qubit on which the filter was applied. The reason is that the working qubit has evolved to identity due to the properties of the filter. The tomographic result was used to design the next filter on the other qubit.  In principle, before applying any filters, it is necessary to reset the two ancilla qubits to $\ket{00}$. As it is difficult to refresh the spins in NMR, an alternative way was adopted in our experiments. For example, to realize $\mathcal{F}_A^2$, we packed it together with $\mathcal{F}_A^0$ and generated a new operator. It can be regarded as a 2-in-1 filter and implemented in the same way. Hence, we avoided the reset operations throughout the experiments and for any individual experiment we just started from the original two-qubit state $\rho_{AB}^{0}$. This feedback-based filtering operations continued to be executed till five filters accomplished and seven 1-qubit tomographies carried out, as shown in Fig.~\ref{scheme}a.

From the above discussions, we have shown that the NMR experiments only contain free J-coupling evolutions and single-qubit unitaries. See the circuit in Fig.~\ref{circuit}. For the J-coupling evolutions, we drove the system to undergo the free Hamiltonian in equation ~\eqref{Hamiltonian} for some time. For local unitaries, we utilized GRadient Ascent Pulse Engineering (GRAPE) techniques~\cite{khaneja2005optimal,ryan2008liquid} to optimize them. The GRAPE approach provided 1 ms pulse width and over 99.8\% fidelity for every local unitary, and furthermore all pulses were rectified via a feedback-control setup in NMR spectrometer to minimize the discrepancies between the ideal and implemented pulses~\cite{Weinstein2004,Moussa2012,Lu2015}.

\textbf{Experimental results and error analysis.} We prepared six input states by varying $\lambda$ from 0.2 to 0.7 with 0.1 step size in the form of equation ~\eqref{inputstate}. After the preparations, we performed two-qubit full state tomography on each state, and reconstructed them as $\rho_0^e$ where the superscript $e$ means experiment. The fidelity between the theoretical state $\rho_{AB}^{0}$ and measured state $\rho_0^e$ is over 98.2\% for each of the six input states. The infidelity can be attributed to the imperfections of PPS, GRAPE pulses and minor decoherence effect. Nevertheless, this infidelity is merely used to evaluate the precision of our input state preparation. For the latter experiments, we only compared the experimental results with $\rho_0^e$, as $\rho_0^e$ was the actual state from which we started the filter-based experiment.

After initial state preparation and each filter, we obtained the reduced density matrix of qubit A and/or B by single-qubit tomography in the subspace where the ancilla qubits are $\ket{00}$ (see Methods). Refer to Fig.~\ref{scheme}a to see the seven gray dots where single-qubit tomography occurred. The average fidelity between the measured single-qubit state and the expected state computed by $\rho_0^e$ is about 99.5\% (Supplementary Table S1), which demonstrates that our filtering operations and single-qubit tomographies are accurate.

\begin{figure}[htb]
\begin{center}
\includegraphics[width= 1\columnwidth]{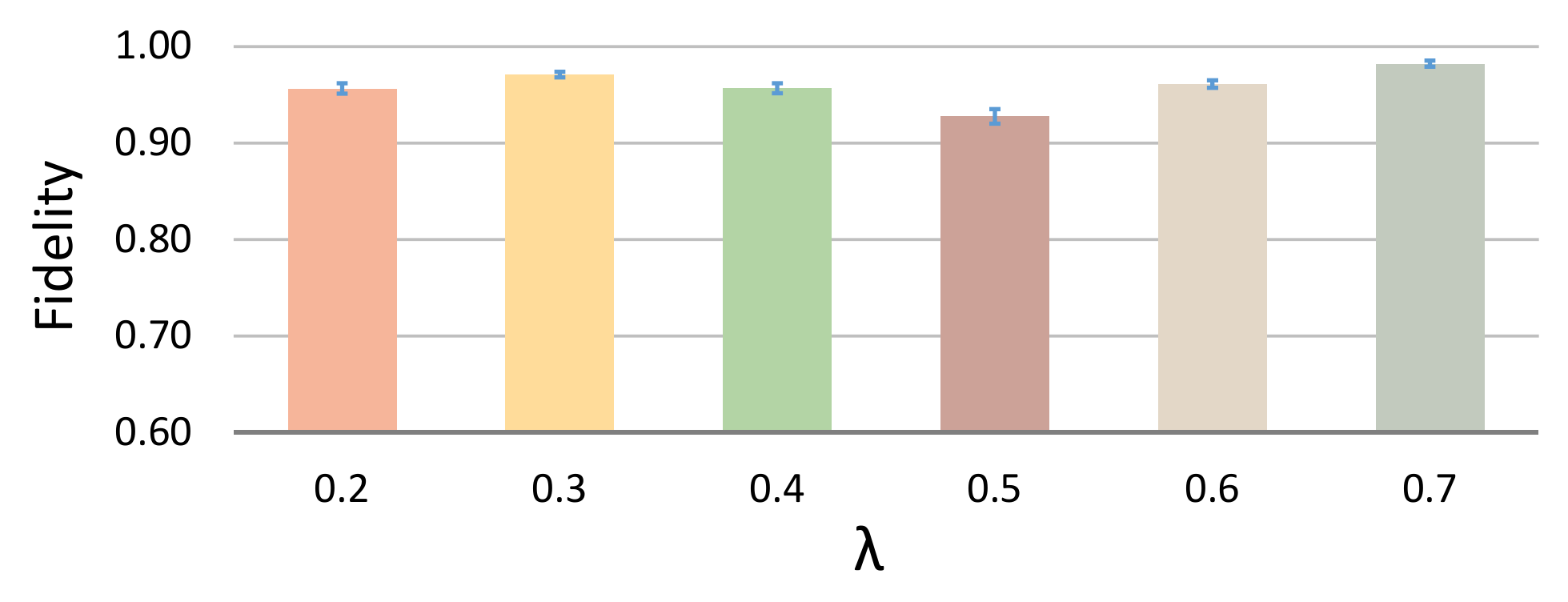}
\end{center}
\setlength{\abovecaptionskip}{-0.00cm}
\caption{\footnotesize{\textbf{Fidelities between $\rho_0^e$ and $\rho_f^e$ for different $\lambda$'s.} $\rho_0^e$ is obtained from two-qubit state tomography right after the creation of the input state $\rho_{AB}^{0}$, and $\rho_f^e$ from the maximum likelihood reproduction of $\rho_{AB}^{0}$ based on the own seven single-qubit states. The error bar comes from the fitting uncertainty when extracting the NMR spectra into quantum states. All fidelities are over 92.0\%, which means the initial two-qubit state is able to be well-reconstructed merely by the seven single-qubit states.  }}\label{result1}
\end{figure}

With the seven single-qubit states in hand, we could reproduce the initially prepared two-qubit state $\rho_0^e$. The maximum likelihood method was adopted here and $\rho_f^e$ was found to be closest to the experimental raw data. Not surprisingly, $\rho_f^e$ is very similar to  $\rho_0^e$, and the fidelity between them for every $\lambda$ is over 92.0\% as illustrated in Fig.~\ref{result1}. Moreover, the real parts of the density matrices $\rho_0^e$ and $\rho_f^e$ are shown in Fig.~\ref{densitymatrix}. The experimental results clearly reveal that the information of the seven single-qubit states collected during the filter-based entanglement distillation procedure already enables the reproduction of the initial two-qubit state. In other words, this filter-based proposal to universally detect and distill entanglement is equivalent compared to doing a two-qubit state tomography.

\begin{figure}[htb]
\begin{center}
\includegraphics[width= 1.1\columnwidth]{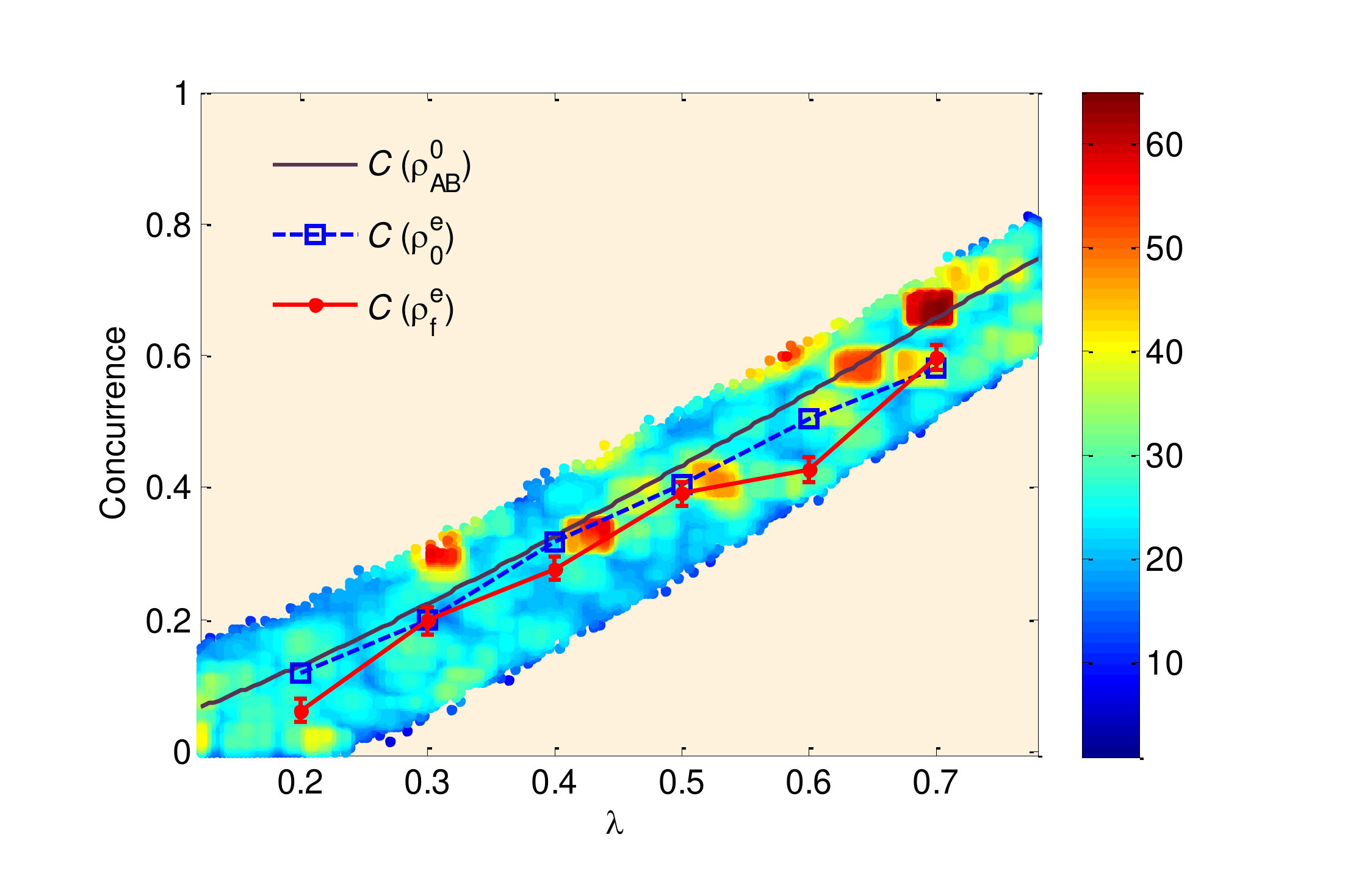}
\end{center}
\setlength{\abovecaptionskip}{-0.00cm}
\caption{\footnotesize{\textbf{Concurrence of $\rho_{AB}^{0}$, $\rho_0^e$, and $\rho_f^e$ as a function of the Bell-state weight $\lambda$.} The brown curve shows the concurrence computed by the theoretical state $\rho_{AB}^{0}$, and exhibits the value of concurrence as a monotonically increasing function of $\lambda$. The blue squares represent the concurrence of $\rho_0^e$, the state obtained from two-qubit state tomography right after the input state preparation. Generally we can roughly assume this state is the truly prepared state, and the following filtering operations are always applied this state as long as we neglect the measurement error of reconstructing $\rho_0^e$. The red circles represent the concurrence of $\rho_f^e$, the state reproduced from the seven single-qubit states. Ideally $\rho_0^e$ and $\rho_f^e$ should be the same if there are no experimental errors. The colored band-region accounts for an artificial noise of the strength 5.79\%, which is roughly estimated from the fitting error 3.00\%, GRAPE imperfection error 1.59\%, and decoherence error 1.20\%. We added this noise on the theoretical state $\rho_{AB}^{0}$, and randomly sampled 2500 states within the noise range for every $\lambda$ (200 values in [0.1, 0.8]). The colored band-region is thus plotted based on the density of projected points out of 2500. }}\label{result2}
\end{figure}

Afterwards, we computed the concurrence for each case with different input two-qubit state. Concurrence is an entanglement monotone defined for a mixed state $\rho$ of two qubits
\be
\mathcal{C}(\rho) = \text{max}\left( 0, \lambda_1-\lambda_2-\lambda_3-\lambda_4\right),
\ee
where $\lambda_1$, $\lambda_2$, $\lambda_3$ and $\lambda_4$ are the eigenvalues of
\be
R = \sqrt{\sqrt{\rho}\left( \sigma_y\otimes\sigma_y \right) \rho^{*} \left( \sigma_y\otimes\sigma_y \right) \sqrt{\rho}}
\ee
in decreasing order. Apparently, the concurrence is proportional to $\lambda$ since $\lambda$ is the weight of Bell-state which is the only term contributing to entanglement. In Fig.~\ref{result2}, the brown curve displays the value of concurrence as a monotonically increasing function of Bell-state weight $\lambda$. The blue squares represent the concurrence of $\rho_0^e$, the state obtained from two-qubit state tomography on the experimentally prepared state. Recall that the preparation fidelity is always over 98.2\% so the blue squares do not deviate much from the brown curve. The red circles represent the concurrence of $\rho_f^e$, which ideally should be the same as blue squares if there are no experimental errors. However, in experiment we have inevitable errors from many factors such as the imprecision of the single-qubit readout stage, the imperfect application of filters and the relaxation, and we need to take them into account.

For convenience, we assume the errors originate from three primary aspects and they are additive. One error is caused by the imprecision of the single-qubit tomography procedure. As we used a least-square fitting algorithm to analyze the outcome spectra and converted the data into quantum states, the fitting induced about 3.00\% uncertainty to the single-qubit readout result. The second is the error from applying imperfect filters in experiment. It mainly comes from the errors of accumulating GRAPE pulses, which is about 1.59\% for each filter operation. The third error, to the lesser extent, is about 1.20\% caused by decoherence. Therefore, in total we estimated at most 5.79\% error might occur in the entire process. We dealt with it as an artificial noise and embedded it into the theoretical input state $\rho_{AB}^{0}$. In simulation, we first discretized $\lambda$ to 200 values from $\lambda=0.1$ to $\lambda=0.8$. For a given $\lambda$, 2500 states were randomly sampled deviated from $\rho_{AB}^{0}$ within 5.79\% noise range. For every sampled state, the concurrence was calculated and projected onto one point in Fig.~\ref{result2}. Hence, a colored band-region was generated considering the density of points. All of our experimental results have fallen into this region, which is consistent with the simulation model.

\begin{figure*}[htb]
\begin{center}
\includegraphics[width= 2\columnwidth]{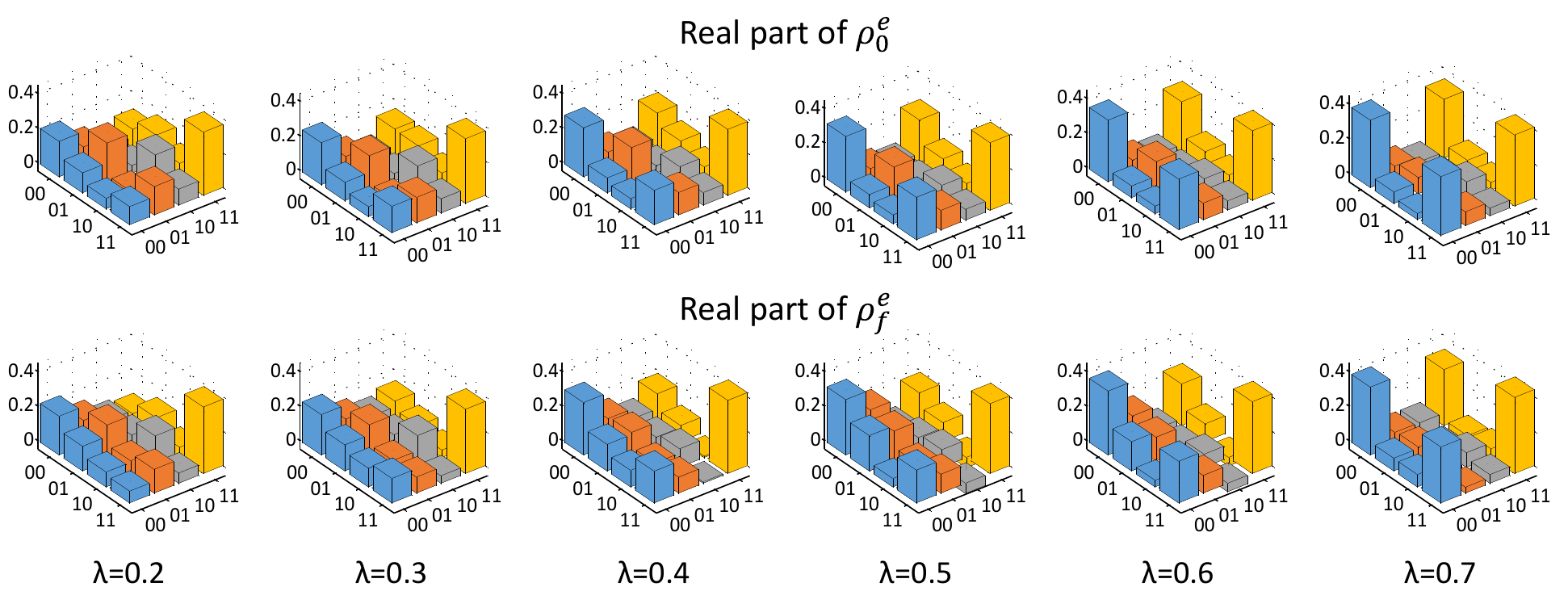}
\end{center}
\setlength{\abovecaptionskip}{-0.00cm}
\caption{\footnotesize{\textbf{Density matrices between $\rho_0^e$ and $\rho_f^e$ for different $\lambda$'s.} Only the real parts are displayed. The upper row shows the density matrices of $\rho_0^e$, which are obtained directly after the input state preparation via two-qubit state tomography. The lower row shows the density matrices of $\rho_f^e$, which are reconstructed through the single-qubit information after implementing the entire circuit in Fig.~\ref{scheme}a. For any $\lambda$, the fidelity between $\rho_0^e$ and $\rho_f^e$ is always above 92\%. }}\label{densitymatrix}
\end{figure*}

\section*{Discussion}

We proved a no-go theorem that there is no way to detect entanglement for an arbitrary bipartite state $\rho_{AB}$ without FST, if only single-copy non-adaptive measurements are allowed. Our observation is due to a nice geometric picture: unless the shape of the
separable states is `cylinder-like', it is not possible to find a projection of the state space to a lower dimensional hyperplane with non-overlapping image for the set of separable states and entangled states. Our method provides a general framework to study the detection of other interesting quantities for a bipartite quantum state, such as positive partial transpose and $k$-symmetric extendibility.

We also have investigated the case of adaptive measurements. It is proposed in~\cite{park2010construction} that
the entanglement measure concurrence for two-qubit states can be determined without FST, via only single-copy measurements.
To implement this protocol, we developed an ancilla-assisted approach to realize the filters. Practically, our technique can be extend to other quantum systems other than optics to implement an amplitude damping channel, which is of great importance in quantum information. By
implementing this protocol, we show that given the experimental data collected for certain state $\rho_{AB}$, this protocol in fact leads to FST of $\rho_{AB}$.
Therefore, this protocol does not lead to universal detection of entanglement of $\rho_{AB}$ without FST.

Our study thus reveals a fundamental relationship between entanglement detection and quantum state tomography. That is, universal detection of entanglement without FST is impossible with only single-copy measurements.
A natural question is what if joint measurements on $r$ copies of the state $\rho_{AB}$ (i.e. $\rho_{AB}^{\otimes r}$) for $r>1$ are allowed. In this case, one indeed can detect entanglement universally for any $\rho_{AB}$ without reconstructing the state, and one example for determining the concurrence of a two-qubit $\rho_{AB}$ is given in~\cite{horodecki2003measuring,augusiak2008universal,bartkiewicz2015method}. However, the protocol of~\cite{augusiak2008universal} involves joint measurements on $4$ copies of $\rho_{AB}$ (i.e. $\rho_{AB}^{\otimes 4}$), which makes the protocol hard to be implemented in practice. It will be interesting to find a smaller $r$ such that joint measurements on c are enough to universally detect the entanglement in $\rho_{AB}$ without full state tomography.

In fact, there are cases that this is possible even for $r=2$. For instance we have found such a scheme that detects the entanglement of an arbitrary two-qubit state $\rho_{AB}$ without FST, if we allow joint measurements on $2$-copies. The idea is that $\rho_{AB}$ is entangled if and
only if ~\cite{augusiak2008universal}
$
\text{Det}(\rho_{AB}^{T_A})<0,
$
where $\rho_{AB}^{T_A}$ is the partial transpose of $\rho_{AB}$ on system $A$. So
we only need to design a scheme with measurements on $\rho_{AB}^{\otimes r}$ which
can give the value of $\text{Det}(\rho_{AB}^{T_A})$. This can indeed be done
without FST (see supplemental materials for details).

Furthermore, if only single-copy measurements are allowed, one cannot determine
the value of $\text{Det}(\rho_{AB}^{T_A})$, even with adaptive measurements. Assume such adaptive measurements exist. Now, we let the input state is maximal mixed state $I/4$, after the measurement, one can compute the determinant. Notice that there exists at least one nonzero traceless $R$ not measured, which means that this measurements can not distinguish between $I/4$ and $I/4+tR$. Therefore, $\text{Det}(I/4+tR^{\Gamma})=\text{Det}(I/4)$ for sufficient small $t$. This then leads to $R=0$.

This strongly supports our no-go results, which indicates that even with adaptive measurements,
universal detection of entanglement with single-copy measurements is impossible without FST.

\section*{Methods}

\textbf{Initialization in NMR.}  We first create the PPS from the thermal equilibrium state, which is a highly mixed state and not yet ready for quantum computation tasks. Since our sample consisting of four $^{13}$C's is a homonuclear system, we simply set the gyromagnetic ratio of $^{13}$C to 1 and write the thermal equilibrium state as
\begin{align}\label{thermal}
\rho_{\text{thermal}}=\frac{{I}}{2^N}+\epsilon\sum _{i=1}^N \sigma_z^i,
\end{align}
where $N=4$ is the number of qubits, ${I}$ is the $2^N\times2^N$ unity matrix, and $\epsilon\approx 10^{-5}$ represents the polarization at room temperature. This initialization step was realized by the spatial average technique~\cite{Cory04031997, dawei2011, 1367-2630-16-5-053015}, and the related pulse sequence is depicted in Fig.~\ref{circuit}. In particular, the gradient pulses represented by Gz crush all coherence in the instantaneous state. The final state after the entire PPS preparation sequence is
\begin{align}\label{pps}
\rho_{0000}=\frac{1-\epsilon}{16}{I}+\epsilon\ket{0000}\bra{0000}.
\end{align}
It is worthy stressing that the large identity does not evolve under any unitary propagator, and it cannot be observed in NMR. Thus we only need to focus on the deviation part $\ket{0000}$ as the entire system behaves exactly the same as it does.

Our aim input state is $\rho_{AB}^0$ in equation (\ref{inputstate}). This mixed state consists of three components: $\ket{\phi_{\mathcal{B}}}$,  $\ket{\phi_{1}}$ and $\ket{\phi_{2}}$ with a weight for each. Typically we repeated every experiment by three times, and created one component in each round as the input. The sequences to prepare all three components are shown in the lower-right inset of Fig.~\ref{circuit}, with all gates applied only on system qubits A and B. (i) For $\ket{\phi_{\mathcal{B}}}$, we applied a Hadamard gate on qubit A, and then a controlled-NOT between A and B; (ii) for $\ket{\phi_{1}}$, we applied $R_x^A(\pi/2)$ and $R_y^B(\pi/2)$; (iii) for $\ket{\phi_{2}}$, we applied $R_y^A(\pi/2)$ and $R_x^B(0.7\pi)$. Subsequently, each component undergoes the whole filtering and measurement procedure respectively, with the final result obtained by summarizing over all three experiments.

\textbf{Single-qubit tomography after each filter.} The entanglement distillation procedure described in Ref.~\cite{park2010construction} involves iterative local filter operations, meaning that every filter depends on the single qubit measurement result before. In experiment we performed single-qubit tomography on system qubit C$_2$ or C$_3$ correspondingly. It requires the measurement of the expectation values of $\sigma_x$, $\sigma_y$  and $\sigma_z$, respectively. In our 4-qubit system, this single-qubit tomography (assuming the measurement of C$_3$) is equivalent to measuring $\ket{0}\bra{0}\otimes \sigma_{x,y,z}\otimes I \otimes \ket{0}\bra{0}$, since we only need to focus in the subspace where the ancilla qubits are $\ket{00}$. To get the expectation values of the observables, a spectrum fitting procedure was utilized to extract these results from NMR spectra. Supplementary Table S1 summarizes all of the single-qubit state fidelities and two-qubit state fidelities for every $\lambda$, and as an example Supplementary Fig. S1 shows the NMR spectra after each filter to measure <$\sigma_x$> and <$\sigma_y$> of the current single qubit (C$_2$ or C$_3$) when $\lambda =0.5$.

\bigskip

{\bf Notes.} After finishing this paper, we noticed a related recent work~\cite{carmeli2015verifying}, where a similar idea
of showing the set of separable states is not `cylinder-like' is developed.

{\bf Author Contributions.} D.L. and T.X. designed and carried out the NMR experiments and simulations; N.Y., Z.J., J.C. and B.Z. made the theoretical proposal and contributed to the analysis of results. X.P. and R.L. supervised the experiment. All authors contributed to the writing of the paper and discussed the experimental procedures and results.

{\bf Acknowledgments.} We thank Y. Zhang for bringing the recent work~\cite{carmeli2015verifying} into our attention. We thank X. Ma for insightful discussions, and are grateful to the following funding sources: NSERC (D.L., N.Y., J.B., B.Z. and R.L.); Industry Canada (R.L.); CIFAR (B.Z. and R.L.); National Natural Science Foundation of China under Grants No. 11175094 and No. 91221205 (T.X. and G.L.), No. 11425523 and  No. 11375167 (X.P.); National Basic Research Program of China under Grant No. 2015CB921002 (T.X. and G.L.).

\clearpage

\onecolumngrid

\begin{widetext}
\center
{\bf Supplementary information: Tomography is necessary for universal entanglement detection
with single-copy observables}
\medskip
\bigskip
\end{widetext}

\onecolumngrid
\appendix

\section{A general method to show that the set of separable states is
  not `cylinder-like'.} We use
two-qubit states as an example, however our method generalized naturally to
the case of any bipartite systems (see supplemental materials for details).

Without loss of generality, we assume there is a set of $S_i$, where $1\leq i\leq 14$, such that there is a function
$g(\tr(S_1\rho_{AB}),\tr(S_2\rho_{AB}),\cdots,\tr(S_{14}\rho_{AB}))=1$ for entangled $\rho_{AB}$, and $0$ otherwise. One can hence view $g$ as an analogue of an entanglement measure which is nonzero if and only if $\rho_{AB}$ is entangled.
Then one can find another observable $R$ such that $\tr R=0$ and $\tr(R^{\dag}S_i)=0$ for any $0\leq i\leq 14$.

Our key observation is that for any non zero traceless $R$ there exists some $\rho_{AB}$ and real $t$ such that $\rho_{AB}+tR$ is non-negative (hence is a quantum state), and $\rho_{AB}$ is separable, $\rho_{AB}+tR$ is entangled. That is, $g(\rho_{AB})$ cannot exist.

To show this, we consider some state on the boundary of separable states, $i.e.$, the two-qubit isotropic states $\rho^{iso}(\alpha)=(1-\alpha)I/4+\alpha\op{\Phi}{\Phi}$ with $\ket{\Phi}=\frac{1}{\sqrt{2}}(\ket{00}+\ket{11})$ being the Bell state. It is known that the isotropic state $\rho^{iso}(\alpha)$ is separable if and only if $\alpha\geq 1/3$~\cite{Horodecki1999,Terhal2000}.

Now we let $\rho_{AB}=\rho^{iso}(\frac{1}{3})=\frac{1}{6}I+\frac{1}{3}\op{\Phi}{\Phi}$,
so $\rho_{AB}$ is separable.
For any $R$, for sufficient small $t$, $\rho_{AB}+tR$ is non-negative since $\rho_{AB}$ is positive(full rank).
Choose $t>0$ such that $\rho_{AB}+tR$ and $\rho_{AB}-tR$ are both non-negative. Therefore, $\rho_{AB}+tR$ and $\rho_{AB}-tR$ are separable.
Notice that $\tr(\sigma\op{\Phi}{\Phi})\leq 1/2$ holds for any separable state $\sigma$~\cite{Vidal1999}, we have
\begin{eqnarray*}
1/2\geq \tr[(\rho_{AB}+tR)\op{\Phi}{\Phi}]\Longrightarrow 0\geq \tr[R\op{\Phi}{\Phi}],\\
1/2\geq \tr[(\rho_{AB}-tR)\op{\Phi}{\Phi}]\Longrightarrow 0\leq \tr[R\op{\Phi}{\Phi}].
\end{eqnarray*}
Thus, $\tr[R\op{\Psi}{\Psi}]=0$ holds for arbitrary maximally entangled state $\ket{\Psi}=(U\otimes I)\ket{\Phi}$. In other words,
\begin{eqnarray*}
R=I\otimes M+N\otimes I,
\end{eqnarray*}
with $\tr(M)=\tr(N)=0$.

Notice that for any non-singular matrix $S_A$ such that $S_ANS_A^{\dag}$ is traceless, $R'=(S_A\otimes I)R(S_A\otimes I)^{\dag}$ also satisfies the property that $\rho_{AB}+tR'$ is separable if and only if $\rho_{AB}$ is separable and $\rho_{AB}+tR'$ is non-negative. According to the previous arguments, we know that $R'$ can be written as $I\otimes M'+N'\otimes I$. Directly, one can conclude that $M=0$. By choosing $S_B$ such that $S_BNS_B^{\dag}$ being traceless, we can obtain that $N=0$. Therefore, $R$ must be $0$. In other words, tomography is required for detecting two-qubit entanglement by using the one copy non-adaptive measurement.

\section{$k$-symmetric extension}
In this section, we will show that the sets of states allowing $k$-symmetric extension are not
‘cylinder like’ for $k\geq 2$, even for two-qubit system, where a state $\rho_{AB}$ is called $k$-symmetric extendable if and only if there exists $\sigma_{AB_1B_2\cdots B_k}$ such that $\sigma_{AB_i}=\rho_{AB}$ for all $1\leq i\leq k$.

We first recall that: A two-qudit Werner state is a state invariant
under the $U\otimes U$ operator for all unitary $U$ and
has the following form
\begin{equation*}
  \rho_W(\psi^{-}) = \frac{1+\psi^{-}}{2} \rho^{+} +
  \frac{1-\psi^{-}}{2} \rho^{-},
\end{equation*}
where $\psi^{-} \in [-1,1]$ is the parameter, $\rho^{+}$ and
$\rho^{-}$ are the states proportional to the projection of the
symmetric subspace and anti-symmetric subspace
respectively.
$\rho_W(\psi^{-})$ is $k$-symmetric extendable iff
$\psi^{-} \ge -(d-1)/k$ proved in~\cite{Johnson2013}.

Assume that the set of states allowing $k$-symmetric extension is cylinder like. In other words there exists some traceless Hermitian operator $R$ such that $\rho+tR$ is $k$-symmetric extendable if $\rho$ is $k$-symmetric extendable and $\rho+tR$ is positive semidefinite.

Choose $\rho_0=\rho_W(-(d-1)/k)$, then $\rho_0>0$ and for any $R$, there exists small $t$ such that $\rho_0+tR$ and $\rho_0-tR$ are all positive semidefinite. Therefore, they are both $k$-symmetric extendable. As a direct consequence, we have Werner states $\sigma_1$ and $\sigma_2$ are both $k$-symmetric extendable,
\begin{eqnarray*}
\sigma_1&=&\int_U (U\otimes U)(\rho_0+tR)(U\otimes U)^{\dag} dU\\
        &=& \rho_0+t \tr(RP^{+})(\rho^{+}-\rho^{-}),\\
\sigma_2&=&\int_U (U\otimes U)(\rho_0-tR)(U\otimes U)^{\dag} dU,\\
        &=& \rho_0-t \tr(RP^{+})(\rho^{+}-\rho^{-}),\\
\end{eqnarray*}
where $P^{+}$ and $P^{-}$ are the projection onto the symmetric subspace and anti-symmetric subspace
respectively.

By using condition of~\cite{Johnson2013}, one can conclude that
\begin{eqnarray*}
\frac{1-(d-1)/k}{2}+t \tr(RP^{+})\geq \frac{1-(d-1)/k}{2},\\
\frac{1-(d-1)/k}{2}-t \tr(RP^{+})\geq \frac{1-(d-1)/k}{2}.
\end{eqnarray*}
Therefore,
\begin{equation*}
\tr(RP^{+})=\tr(R\rho^{+})=\tr(RP^{-})=0.
\end{equation*}
Similar technique can be applied for $(V\otimes I)\rho (V\otimes I)^{\dag}$ with unitary $V$. That leads us to
\begin{equation*}
\tr(R(V\otimes I)\rho^{+} (V\otimes I)^{\dag})=0.
\end{equation*}
That is, $R^{\Gamma}$, the partial transpose of $R$, is orthogonal to all maximally entangled state $(V\otimes I)\ket{\Phi}$,
\begin{eqnarray*}
&&\tr(R^{\Gamma}(V\otimes I)\op{\Phi}{\Phi}(V\otimes I)^{\dag})\\
&=&\tr(R(V\otimes I)(P^{+}-P^{-})(V\otimes I)^{\dag})\\
&=&0.
\end{eqnarray*}
Now we write $R$ as follows,
\begin{eqnarray*}
R^{\Gamma}_{AB}=I\otimes M+N\otimes I+X_{AB},
\end{eqnarray*}
with $\tr(M)=\tr(N)=0$, and $X_{A}=X_{B}=0$. This can be done by simply choosing $M=\frac{(R^{\Gamma}_{AB})_B}{d}$, $N=\frac{(R^{\Gamma}_{AB})_A}{d}$ and $X_{AB}=R^{\Gamma}_{AB}-I\otimes \frac{(R^{\Gamma}_{AB})_B}{d}-\frac{(R^{\Gamma}_{AB})_A}{d}\otimes I$.

For sufficient small $s$, $I_{AB}/d+sX_{AB}$ is a choi matrix of some unital quantum channel.
According to Theorem 1 of \cite{Mendl2009}, we know that $I_{AB}/d+sX_{AB}$ is a linear combination of the density matrix of maximally entangled states, so is $X_{AB}$. Thus,
\begin{eqnarray*}
0=\tr(R^{\Gamma}_{AB}X_{AB}^{\dag})=\tr(X_{AB}X_{AB}^{\dag}),
\end{eqnarray*}
where we use the fact that $I\otimes M+N\otimes I$ is orthogonal to all maximally entangled states.
Thus, $X_{AB}=0$ and $R$ can be written as
\begin{eqnarray*}
R^{\Gamma}_{AB}=I\otimes M+N\otimes I.
\end{eqnarray*}

Now, notice that for any non-singular matrix $S_A$ such that $S_ANS_A^{\dag}$ is traceless, $R'=(S_A\otimes I)R(S_A\otimes I)^{\dag}$ also satisfies that $\rho_{AB}+tR'$ is $k$-symmetric extendable if and only if $\rho_{AB}$ is $k$-symmetric extendable and $\rho_{AB}+tR'$ is non-negative. Then we know that $R'$ can be written as $I\otimes M'+N'\otimes I$, too. Directly, one can conclude that $M=0$.
Thus, we only need to deal with
\begin{eqnarray*}
R=N^{\Gamma}\otimes I.
\end{eqnarray*}
In the following, we deal with the two-qubit case. Note that there exists local unitary which transforms $R$ into diagonal version $R=Z\otimes I=\diag\{1,1,-1,-1\}$.

We only construct some state $\rho$ which is not 2-symmetric extendable~\cite{Chen2014} and $\rho+R$ is separable. Then such $\rho+R$ is $k$-symmetric extendable for all $k$ while $\rho$ is not 2-symmetric extendable.
\begin{eqnarray}
\rho:&=&\left(\begin{array}{cccc}
x & 0 & 0 & \sqrt{xw}\\
0 & y & \sqrt{yz} & 0\\
0 & \sqrt{yz} & z & 0\\
\sqrt{xw} & 0 & 0 & w
\end{array}\right),\\
\rho+R:&=&\left(\begin{array}{cccc}
x+1 & 0 & 0 & \sqrt{xw}\\
0 & y+1 & \sqrt{yz} & 0\\
0 & \sqrt{yz} & z-1 & 0\\
\sqrt{xw} & 0 & 0 & w-1
\end{array}\right)
\end{eqnarray}
By using the condition of~\cite{Chen2014}, $\rho$ is not 2-symmetric extendable iff
\begin{eqnarray*}
(x+z)^2+(y+w)^2&<& x^2+y^2+z^2+w^2+2xw+2yz,\\
\Leftrightarrow  (x-y)(w-z)&>&0.
\end{eqnarray*}
$\rho+R$ is separable iff
\begin{eqnarray*}
&&(x+1)(w-1)\geq xw,yz,\\
&& (y+1)(z-1)\geq xw,yz.
\end{eqnarray*}

It is direct to see that one can choose some $w>z>x>y>0$ such that
\begin{eqnarray*}
&&(x+1)(w-1)\geq xw\geq yz,\\
&& (y+1)(z-1)\geq xw \geq yz.
\end{eqnarray*}
Actually, we can choose $\epsilon$ to be sufficient small, and
\begin{eqnarray*}
x=y+\epsilon,\\
z=y+2,\\
w=y+2+\epsilon.
\end{eqnarray*}
Therefore, for all nonzero $R$, one can always find $\rho$ and $\rho+R$ such that $\rho+R$ is separable and $\rho$ is not 2-symmetric extendable. This shows that the set of states allowing $k$-symmetric extension is also not `cylinder-like', even for two-qubit system.

\section{The case of joint measurements}

We discuss the case of joint measurements on $r$ copies of $\rho_{AB}$
(i.e. $\rho_{AB}^{\otimes r}$) with $r>1$. We take the two-qubit case as
an example. In this case, it is known that $\rho_{AB}$ is entangled if and
only if ~\cite{augusiak2008universal2}
\begin{equation}
\text{Det}(\rho_{AB}^{T_A})<0,
\end{equation}
where $\rho_{AB}^{T_A}$ is the partial transpose of $\rho_{AB}$ on system $A$.

Notice that the determinant of $\rho_{AB}^{T_A}$ is a polynomial of degree $4$
in terms of the matrix entries of $\rho_{AB}$, so it can be detected by measuring
only a single observable on $4$ copies of $\rho_{AB}$ (i.e. $\rho_{AB}^{\otimes 4}$)~\cite{augusiak2008universal2}.

For the case of joint measurements with $r=2$, however, one cannot measure
only a single observable to get $\text{Det}(\rho_{AB}^{T_A})$. Nevertheless, we
show that the value of $\text{Det}(\rho_{AB}^{T_A})$ can be get without full state tomography with only a single joint measurement on $\rho_{AB}^{\otimes 2}$.

To see how this works, we rewrite $\rho_{AB}^{T_A}$ as
\[
\rho_{AB}^{T_A}=\begin{pmatrix}
R & S\\
S^{\dag} & T
\end{pmatrix},
\]
with $R$ is a $3\times 3$ matrix, $S$ is $3\times 1$,
and $T$ is $1\times 1$.

We can first determine
$R$ by single-copy measurements on $\rho_{AB}^{T_A}$,
where $9$ independent observables need to be measured.
After that, $T$ is known by the normalization condition $\tr\rho_{AB}^{T_A}=1$
This does not lead to a full state tomography on $\rho_{AB}$,
since $S$ is undetermined, with $6$ free real parameters.

However, after knowing $R$ and $T$, $\text{Det}(\rho_{AB}^{T_A})$
is a polynomial of degree $2$
in terms of the matrix entries of $\rho_{AB}$, so it can be detected by measuring
only a single observable on $2$ copies of $\rho_{AB}$ (i.e. $\rho_{AB}^{\otimes 2}$).
Together with the measurements on $R$ and $S$, we have total $10$ measurement
outcomes that determine universally whether $\rho_{AB}$ is entangled or not. And
this does not lead to full state tomography of $\rho_{AB}$, since $\rho_{AB}$ needs
$15$ real parameters to determine.

In this way, we determine whether $\rho_{AB}$ is entangled or not without full state tomography, by measuring a single observable on $\rho_{AB}^{\otimes 2}$ together
with single copy measurements. This indicates that our no-go results for single copy measurements fail in the case if joint measurements are allowed, even for $r=2$.

\section{Experimental results of single-qubit tomography after each filter}
Table \ref{single_qubit_fidelity} summarizes all of the single-qubit state fidelities and two-qubit state fidelities for every $\lambda$, and as an example Fig. \ref{single_qubit_spec} shows the NMR spectra after each filter to measure <$\sigma_x$> and <$\sigma_y$> of the current single qubit (C$_2$ or C$_3$) when $\lambda =0.5$.

\begin{table}[htb]
\begin{center}
\includegraphics[width= 0.7\columnwidth]{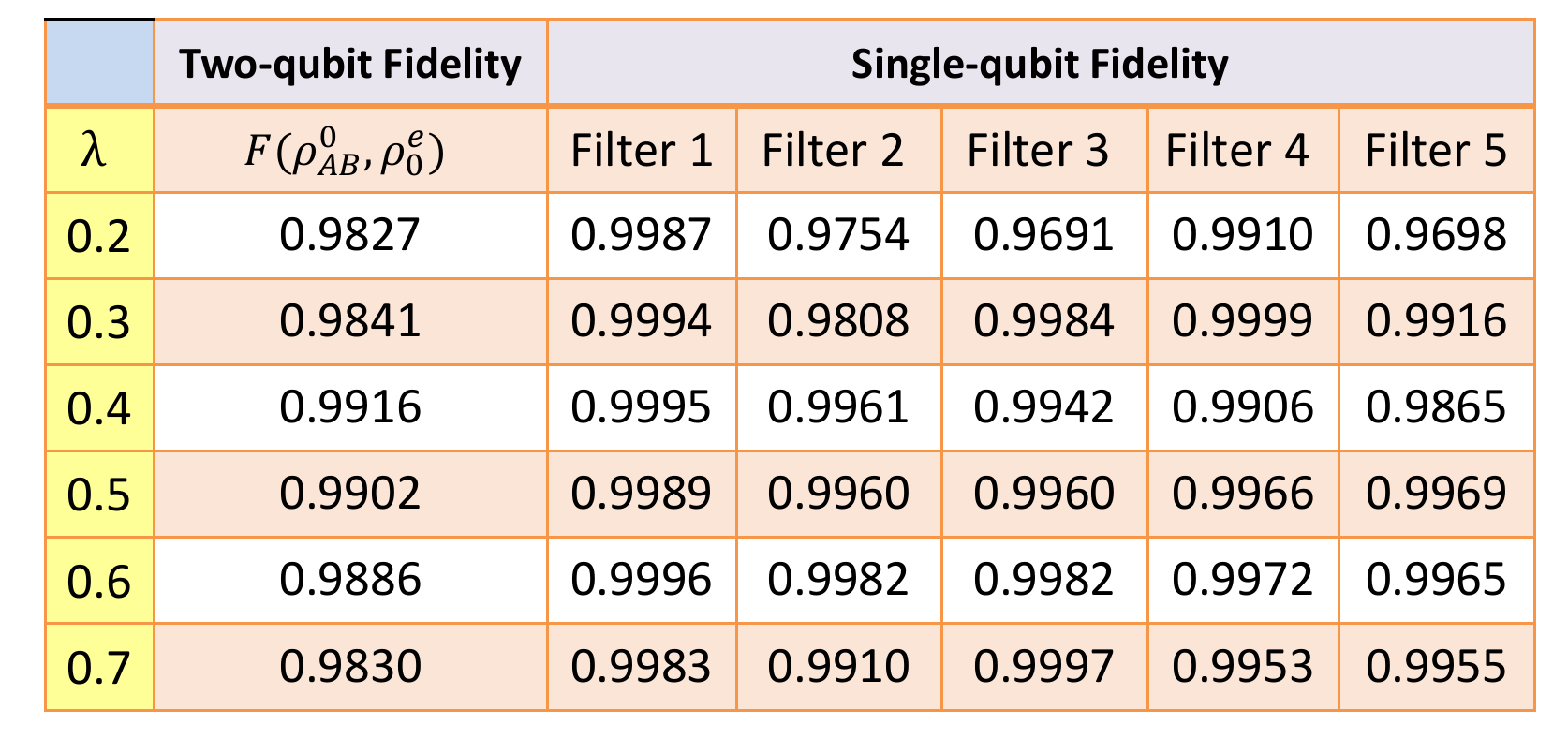}
\end{center}
\setlength{\abovecaptionskip}{-0.00cm}
\makeatletter
\renewcommand{\thetable}{S\@arabic\c@table}
\makeatother
\caption{\footnotesize{Fidelities of the experimental  results compared with the theoretical ones for every $\lambda$. $F(\rho_{AB}^0, \rho_0^e)$ is the fidelity between the theoretical 2-qubit state $\rho_{AB}^0$ and $\rho_0^e$ which is the truly prepared 2-qubit state. Meanwhile, the fidelities of the five single-qubit states after each filter are also shown.}}\label{single_qubit_fidelity}
\end{table}

\begin{figure}[htb]
\begin{center}
\includegraphics[width= 0.7\columnwidth]{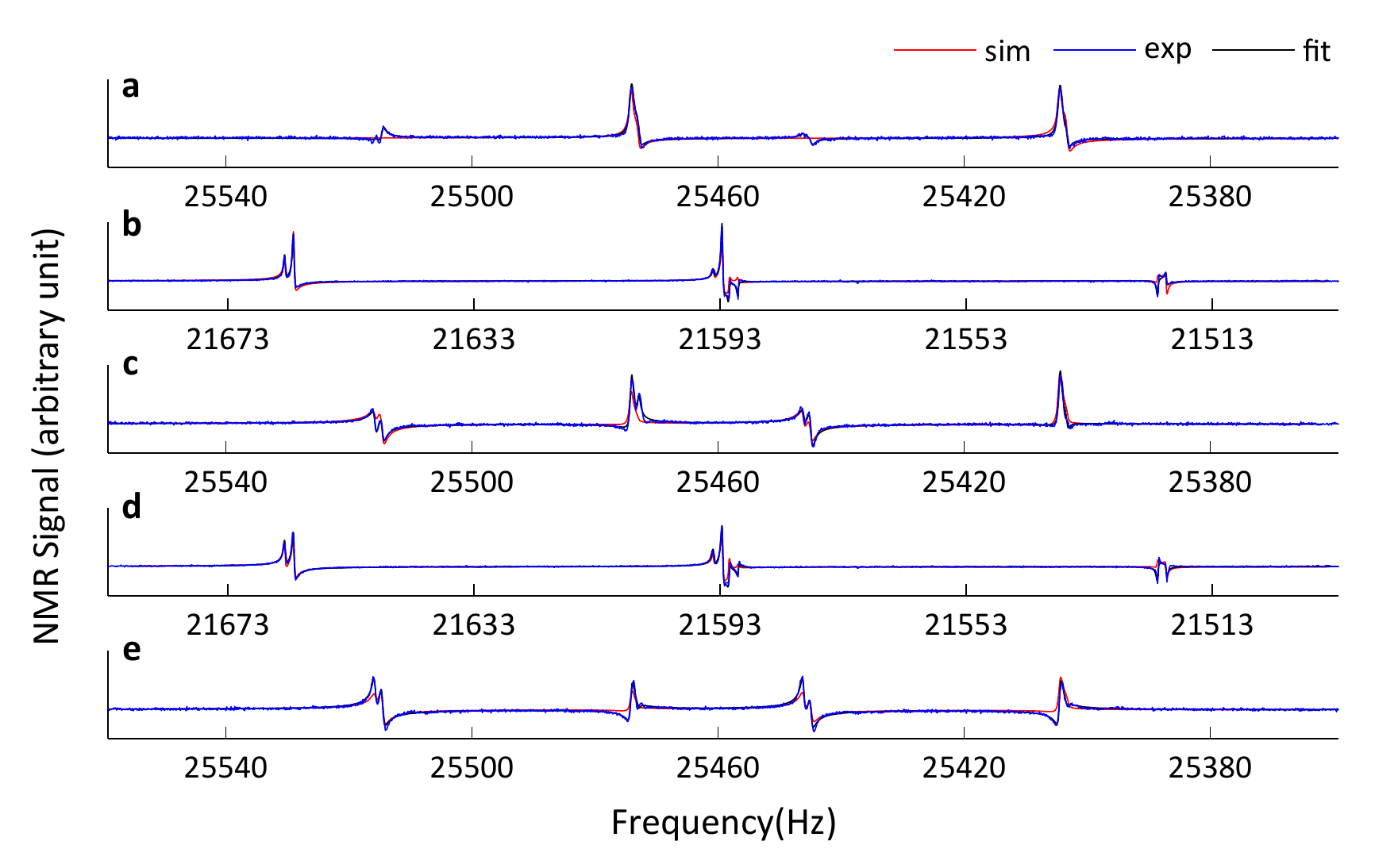}
\end{center}
\setcounter{figure}{0}
\makeatletter
\renewcommand{\thefigure}{S\@arabic\c@figure}
\makeatother
\setlength{\abovecaptionskip}{-0.00cm}
\caption{\footnotesize{NMR spectra of measuring <$\sigma_x$> and <$\sigma_y$> of single qubit after each filter for $\lambda =0.5$.  (a-e) are produced after filter 1-5, respectively. (a, c, e) show spectra of C$_2$, and (b, d) show C$_3$. The red curve is the simulation result assuming the input state is $\ket{0}\bra{0}\otimes \rho_0^e \otimes \ket{0}\bra{0}$,  and the blue curve is the experimental result which can be used to extract <$\sigma_x$> and <$\sigma_y$> of the current qubit. The black curve shows the fitting spectrum to obtain <$\sigma_x$> and <$\sigma_y$>. To measure <$\sigma_z$>, we rotated it to $\sigma_x$ with a $\pi/2$ pulse around $y$-axis and then measured. It can be seen that the fitting matches extremely well with the experimental result, which means our readout values are very accurate.}}\label{single_qubit_spec}
\end{figure}

\bibliographystyle{unsrt}

\end{document}